\documentstyle[epsfig,psfig]{texas}

\begin{document}

\title{The First Light seen in the redshifted 21--cm radiation}

\author{Paolo Tozzi$^{1,2}$, Piero Madau$^{1,3}$, 
Avery Meiksin$^{4}$, and Martin J. Rees$^{3}$} 

\address{(1) Space Telescope Science Institute \\
3700 San Martin Drive,
Baltimore MD 21218 \\
}  
\address{(2)Department of Physics and Astronomy\\
 The Johns Hopkins 
University, Baltimore, MD 21218\\
} 
\address{(3)Institute of Astronomy\\
 Madingley Road, Cambridge CB3 0HA, UK  \\
} 
\address{(4)Institute for Astronomy, University of Edinburgh \\
Royal Observatory, Edinburgh EH9 3HJ, UK\\
}

\begin{abstract}
We show how the investigation of the redshifted 21--cm radiation can
give insight into the development of structures in the early universe
(at redshifts $z>5$).  In particular we investigate: the epoch of the
first light; the fluctuations in the redshifted 21--cm emission induced
by the density inhomogeneities in CDM dominated universes; the
emission and absorption shells that are generated around the first
bright quasars.  Such features can be observed with the next
generation radio facilities.  
\end{abstract}

\section{The general framework}

The diffuse Intergalactic Medium (IGM) at very high redshift (between
recombination and full reionization at $z>5$) can be observed in the
redshifted 21--cm radiation against the cosmic background.  The signal
can be detected in emission or in absorption depending on whether the spin
temperature $T_S$ is larger or smaller than the cosmic background
temperature $T_{CMB}=2.73\,  (1+z)$.

This can happen if $T_S$ is coupled via collisions to the kinetic
temperature $T_K$ of the IGM.  However, the density contrast on Mpc
scales at very early epochs is so low that the collision coupling is
inefficient, and therefore the IGM is expected to be invisible against
the CMB (Madau, Meiksin \& Rees, 1998 \cite{ref1}, hereafter MMR).  On
the other hand, large, massive regions at high density contrast are
extremely rare in most of the hierarchical CDM universes at such high
redshifts.

There is a however another mechanism that makes the diffuse hydrogen
visible in the redshifted 21cm line: the Wouthuysen-Field effect.  In
this process, a Ly$\alpha$ photon field mixes the hyperfine levels of
neutral hydrogen in its ground state via intermediate transition to
the $2p$ state.  A detailed picture of the Wouthuysen-Field effect can
be found in Meiksin (1999, \cite{ref2}) and Tozzi, Madau, Meiksin \&
Rees (1999, \cite{ref3}, hereafter TMMR).  The process effectively
couples $T_S$ to the color temperature $T_\alpha$ of a given
Ly$\alpha$ radiation field (Field 1958, \cite{ref4}).  The color
temperature is easily driven toward the kinetic temperature $T_K$ of
the diffuse IGM due to the large cross section for resonant scattering
(Field 1959, \cite{ref5}).  In this case the spin temperature is:
\begin{equation}
T_S=\frac{T_{\rm CMB}+y_\alpha T_K}{1+y_\alpha}\, ,
\end{equation}
where $y_\alpha\simeq 3.6\, 10^{13}P_\alpha/T_K$, and $P_\alpha$ is
the total rate at which Ly$\alpha$ photons are scattered by an hydrogen
atom.  

However, the same Ly$\alpha$ photon field also re--heats the diffuse
gas, driving $T_K$ toward larger values.  The thermal history of the
diffuse IGM then results from the competition between adiabatic
cooling due to the cosmic expansion and re--heating due to the photon
field.  In the absence of a contribution from a strong X-ray background,
the thermal history of the IGM can be written simply as:
\begin{equation}
{{dT_K}\over{dz}}={{2\mu}\over 3}{\dot E\over k_B}
{{dt}\over{dz}} + 2{{T_K}\over{(1+z)}},
\end{equation}
where $\dot E$ is the heating rate due to recoil of scattered
Ly$\alpha$ photons.  Here $\mu=16/13$ is the mean molecular weight
for a neutral gas with a fractional abundance by mass of hydrogen
equal to 0.75.

Prior to the generation of the photon field, the IGM is neutral and
cold, at a temperature $T_K\simeq 2.6\, 10^{-2} (1+z)^2$ (Couchman
1985, \cite{ref6}) given only by the adiabatic cooling after
recombination.  At the onset of the re--heating sources, there will be
coupling between the kinetic and the spin temperature.  An observation
at the frequency $1420/(1+z)$ MHz will detect absorption or emission
against the CMB, with a variation in brightness temperature with
respect to the CMB value:
\begin{equation}
\Delta T_b\simeq  (2.9\,{\rm mK})\, h^{-1} \,\eta 
\Big({{\Omega_b\, h^2}\over{0.02}}\Big) {(1+z)^2 
\over [\Omega_M(1+z)^3+\Omega_K(1+z)^2+\Omega_{\Lambda}]
^{1/2} }\, ,
\end{equation}
where $\Omega_K=1-\Omega_M-\Omega_{\Lambda}$ is the curvature
contribution to the present density parameter, $\Omega_\Lambda$ is the
cosmological constant, $\Omega_b$ is the baryon density, and $\eta
\equiv (T_{CMB}-T_S)/T_S$.  

Observations of such variation in the brightness temperature can be
used to investigate the thermal history of the IGM, and the underlying
birth and evolution of the radiation sources.  All the following
results are presented and discussed in TMMR (\cite{ref3}).

\section{The epoch of the First Light}

We first investigate a simple situation in which, at a given redshift
$z_{th}$, the Ly$\alpha$ photon field reaches a thermalization rate $
P_{th} \approx 7.6\times 10^{-13} ~{\rm s}^{-1}~ (1+z)$; for such a
value $T_S$ is driven effectively toward $T_K$ (see MMR, \cite{ref1}).
If the IGM is heated only by the same Ly$\alpha$ photons, there will
be a transient epoch where $T_S<T_K$, (i.e., $\eta <0$) and an
absorption feature appears at the corresponding redshifted frequency.
This effect necessarily has a limited extension in time and thus in
frequency space, since $T_S$ becomes larger than $T_{CMB}$ on a
relatively short timescale.  However, the signal is easily detectable
with a resolution of few MHz, and, most of all, has a large effect
since $|\eta| >>1$ when $T_S<<T_{CMB}$.  Such a strong feature marks
the transition from a cold and dark universe, to a universe populated
with radiation sources.
\begin{figure}
\centering
\psfig{file=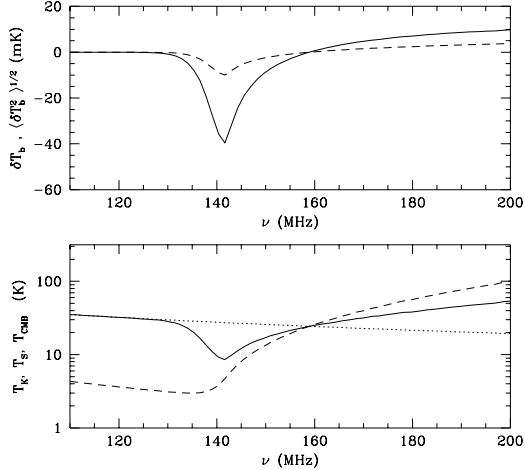,width=8cm}
\caption{\sl Top panel: $\Delta T_b$ for a resolution of 1' and 1 MHz
in frequency, both for the fluctuations (dashed) and for the
continuous distribution of the IGM (continuous line), assuming
$P_{\alpha}$ to be equal to the thermalization rate at $z_{th}\simeq
9$ in a critical universe.  Bottom panel: the corresponding $T_S$
(continuous) and $T_K$ (dashed) are shown together with $T_{CMB}$
(dotted line).}
\label{fig1}
\end{figure}

If we assume that the Ly$\alpha$ field reaches the thermalization value
when $z_{th}=9$ on a timescale $\tau\simeq 10$ Myrs, the IGM will be
visible in absorption for $\simeq 10\div 30$ Myrs, corresponding to
$\Delta T_b \simeq 40$ mK over a range of $\simeq 5$ MHz.  In the top
panel of figure \ref{fig1} such a signature is shown as a function of the
observed frequency.  In the bottom panel the corresponding thermal
evolution for the IGM is shown.

\begin{figure}
\centering
\psfig{file=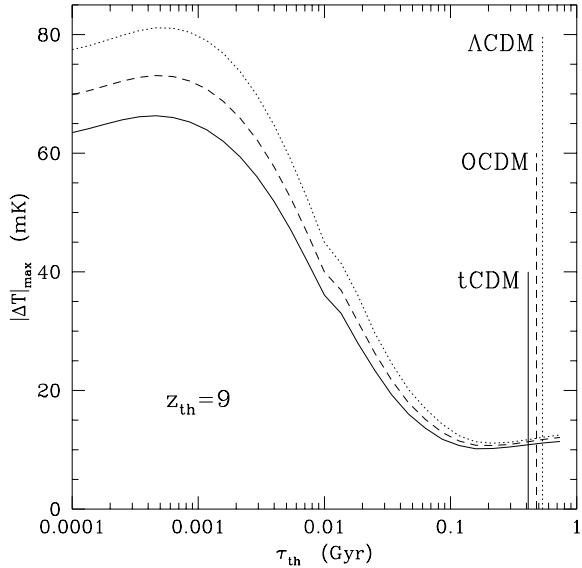,width=8cm}
\caption{The maximum of the absorption is shown as a function of the
timescale on which the Ly$\alpha$ photon field reaches (exponentially)
the thermalization value at the same $z_{th}=9$.  The different curves
corresponds to $\Omega_0=1$, $h=0.5$ (continuous line),
$\Omega_0=0.4$, $h=0.65$ (dashed line), and $\Omega_0=0.3$, $\Lambda =
0.7$, $h=0.7$ (dotted line).  The vertical lines mark the age of the
universe at $z=9$ in the three cases.  }
\label{fig2}
\end{figure}
 
Such results are weakly dependent on the epoch $z_{th}$ and on the
adopted cosmology.  However, the amplitude of the detected signal will
be strongly dependent on the timescale $\tau$ on which the Ly$\alpha$
field reaches the thermalization value.  In figure \ref{fig2} the
maximum of the absorption is plotted for different timescales $\tau$
in three representative cosmologies.  The signal is always larger than
10 mK; note however that for $\tau> 30$ Myr, the absorption is spread
out over a large interval in frequency, especially at $\nu < 100$ MHz
where the sensitivity of radio telescopes becomes lower (see
TMMR).

\section{The density field}

After re--heating and before reionization, $T_S >> T_{CMB}$ holds, and
the IGM is detectable only in emission.  However, $\eta \leq 1$
always, and the effect due to the continuum distribution of a diffuse
IGM is not as strong as in the absorption case.  Such a small positive
offset with respect to the CMB background can be difficult to detect.
On the other hand, fluctuations in the redshifted 21--cm emission, which
reflect fluctuations in the density of the IGM, are
at least two orders of magnitude larger than the intrinsic CMB
fluctuations on scales of $\simeq 1 $ arcmin.

\begin{figure}
\centering
\psfig{file=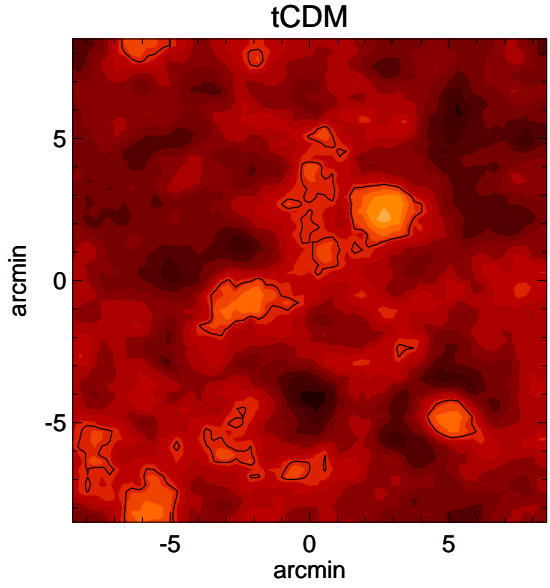,width=9cm}
\medskip
\medskip
\medskip
\medskip
\caption{Radio map of redshifted 21--cm emission against the CMB
in a tCDM cosmology at $z=8.5$.  The linear size of the box is $20h^{-1}$ 
(comoving) Mpc. The point spread function of the synthesized beam 
is assumed to be a spherical top--hat with a width of 2 arcmin. The 
frequency window is 1 MHz around a central frequency of $150$ MHz.
The color intensity goes from $1$ to $6\,$ $\mu$Jy per beam.
For clarity, the contour levels outline regions with
signal greater than $4\,\mu$Jy per beam. 
}
\label{fig3}
\end{figure}

\begin{figure}
\centering
\psfig{file=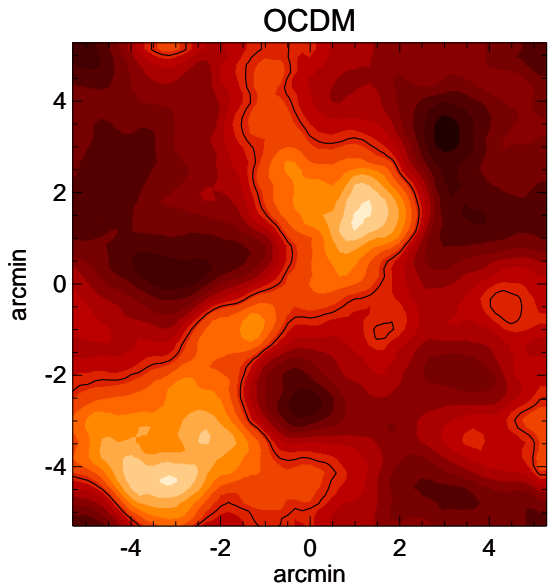,width=9cm}
\medskip
\medskip
\medskip
\medskip
\caption{Same as figure \ref{fig3} for OCDM.
\label{fig4}}
\end{figure}

These fluctuations correspond to scales of a few comoving Mpc, and are
in the linear regime at $z>5$.  In this case the fluctuations induced
in the brightness temperature will be directly proportional to $\Delta
\rho /\rho$, allowing a straightforward reconstruction of the
perturbation field at that epoch.  In figure \ref{fig3} and \ref{fig4}
we show results for two cosmologies, a tilted CDM universe with
critical density (tCDM), and an open $\Omega_0=0.4$ universe (OCDM).
In both cases the fluctuations are normalized to reproduce the local
abundance of clusters of galaxies.  In OCDM the fluctuations are much
larger (a factor of 3) since the evolution of the perturbation is
strongly suppressed in an open universe with respect to the critical
case, and for a given local normalization, the amplitude of the
perturbations at high $z$ is correspondingly larger.  In both figures
the density field has been evolved with a collisionless N--body
simulation of 64$^3$ particles using the Hydra code (Couchman, Thomas,
\& Pearce 1995, \cite{ref7}).  The box size is $20 h^{-1}$ comoving
Mpc, corresponding to 17 (11) arcmin in tCDM (OCDM). The baryons are
assumed to trace the dark matter distribution without any biasing.
Since the level of fluctuations ranges from a few to $\simeq 10$
$\mu$Jy per beam (with a resolution of 2 arcmin), it seems possible
that observations with the {\it Square Kilometer Array} (Braun 1998,
\cite{ref8}, see also http://www.nfra.nl.skai) may be used to
reconstruct the matter density field at redshifts between the epoch
probed by galaxy surveys and recombination, on scales as small as
$0.5-2$ $h^{-1}$ comoving Mpc, i.e. masses in the range between
$10^{12}$ and $10^{13}\; h^{-1} M_\odot$.

\section{The first quasars}

\begin{figure}
\centering
\psfig{file=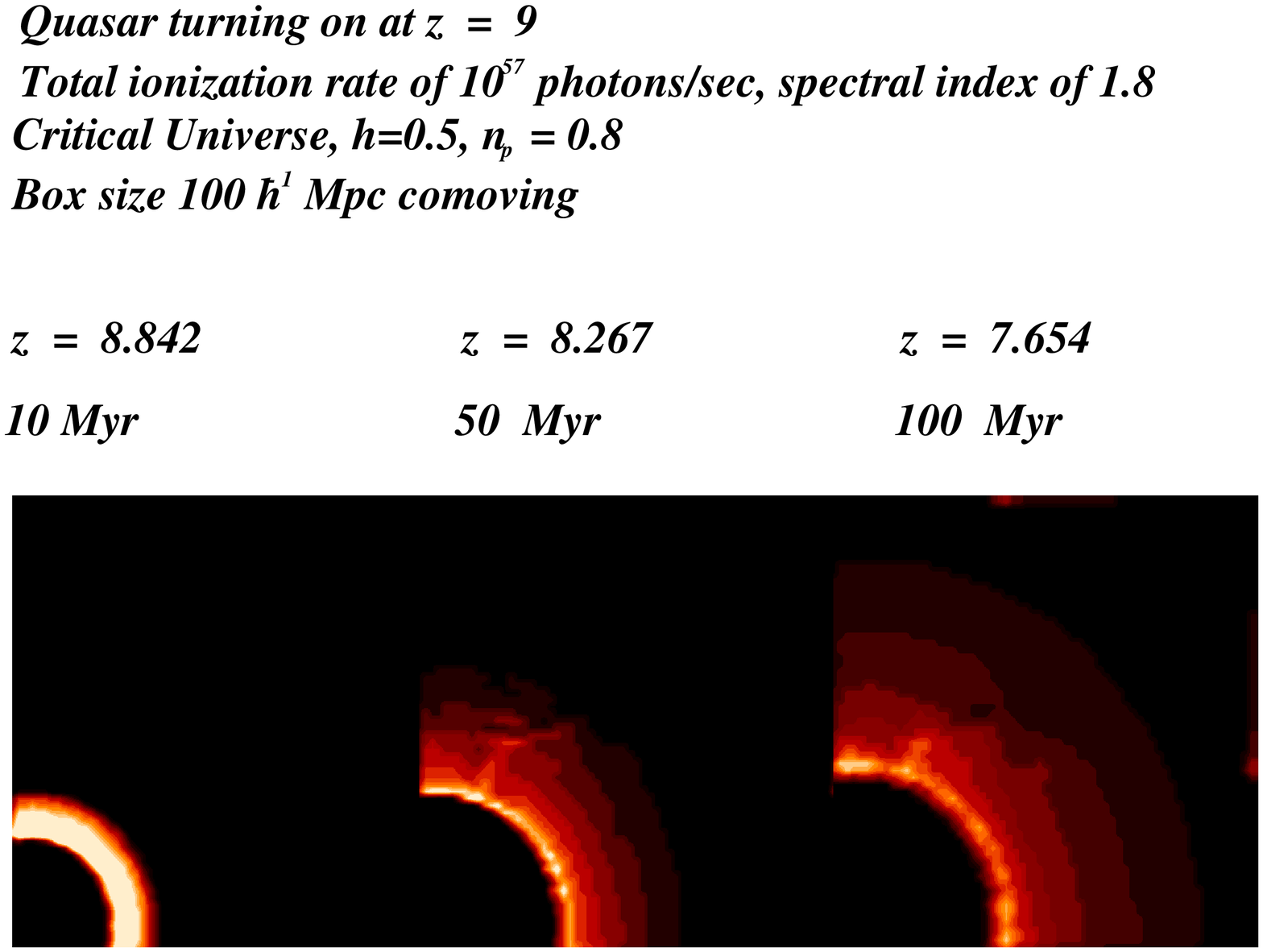,width=15cm}
\caption{21--cm emission against the CMB from the region surrounding a
quasar source (lower left corner, in the center of an HII zone),
revealed once the IGM is heated above the CMB by soft X-rays from the
quasar.  The angular resolution is $2$ arcmin, the frequency depth is
1MHz, and the color levels range from 0 to 10 mK with respect to the
CMB level (dark).  The temperature of the IGM is assumed to be
$T_K=T_{CMB}$ at large distances from the quasar.
\label{fig5}}
\end{figure}

If re--heating is provided by a single quasar (without any other
source of radiation), 21--cm emission on Mpc scales will be produced
in the quasar neighborhood (outside the HII bubble) as the medium
surrounding it is heated to $T_S=T_K>T_{\rm CMB}$ by soft X-rays from
the quasar itself.  The size and intensity of the detectable 21--cm
region will depend on the quasar luminosity and age.  In particular the
intensity in the emission weakens with radius and with the age of the
quasar.

We calculated the kinetic temperature around a typical quasar, along
with the neutral IGM fraction and Ly$\alpha$ flux.  The resulting
radial temperature profiles were then superimposed on the surrounding
density fluctuations as computed using Hydra.  In figure \ref{fig5} a
sequence of snapshots after 10, 50 and 100 Myr after the birth of a
quasar at $z=8.5$ are shown in a box of 100$h^{-1}$ Mpc
(comoving). The visual effect is due to the convolution of the spin
temperature profile with the (linearly) perturbed density field around
the quasar.  The temperature of the IGM at great distances from the
quasar is assumed to be $T_K\simeq T_{CMB}$, and the signal goes to zero.  
In the figure the color ranges from 0 to $10$ mK with
respect to the CMB level, which is black.

\begin{figure}
\centering
\psfig{file=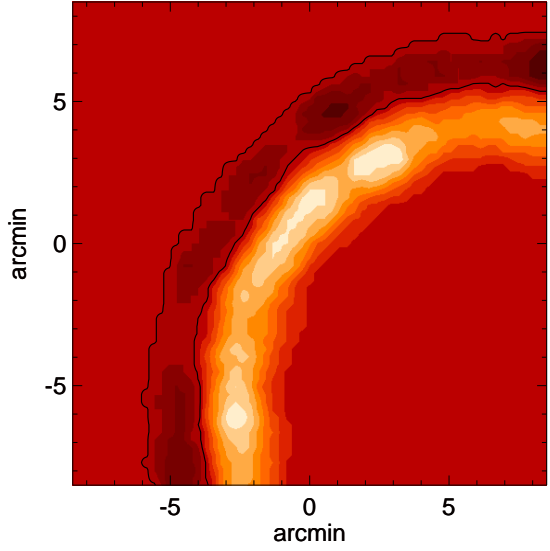,width=9cm}
\medskip
\medskip
\medskip
\medskip
\caption{21--cm emission and absorption against the CMB from the
region surrounding a quasar source (lower right corner, in the center
of an HII zone), revealed once the IGM is heated above the CMB by soft
X-rays from the quasar.  The linear size of the box is of $20h^{-1}$
comoving Mpc (tCDM), the angular resolution is $2$ arcmin, the
frequency depth 1 MHz.  The color levels range from $-3$ $\mu$Jy to
$3$ $\mu$Jy per beam. The contours mark the absorption ring. The
quasar turns on at $z = 8.5$ with a ionizing photon luminosity of
$10^{57}$ photons s$^{-1}$, and is observed after $10$ Myr.  The
temperature of the IGM beyond the light radius is assumed to be
$T_K\simeq 2.6 \, 10^{-2} (1+z)^2\simeq 2.6$ K $< T_{CMB}$.
\label{fig6}}
\end{figure}

\begin{figure}
\centering
\psfig{file=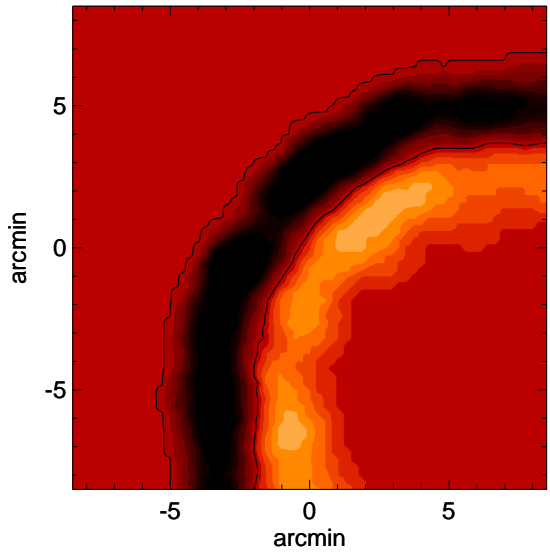,width=9cm}
\medskip
\medskip
\medskip
\medskip
\caption{Same as figure \ref{fig6}, but the spectrum of the quasar has an
exponential cutoff at energies larger than the Lyman limit.  The
absorption ring is much more evident and deep.  
\label{fig7}}
\end{figure}

Another situation occurs when the temperature of the IGM at large
distances from the quasar is lower than $T_{CMB}$, e.g., $T_K=2.6\,
10^{-2}(1+z)^2$ K.  In this case the emission region is followed by an
absorption ring, since the Ly$\alpha$ photons reach regions where
$T_K<T_{CMB}$.  The radio map resulting from a quasar `sphere of
influence' 10 Myr after it turns on at $z=8.5$ (tCDM) is shown in
figure \ref{fig6}.  The signal ranges from about $-3$ $\mu$Jy to $3
\mu$Jy per beam (with a 2 arcmin resolution).  The absorption region
is limited to a very sharp edge.  However, in figure \ref{fig7} we
show a quasar with the same Ly$\alpha$ luminosity but with an
intrinsic exponentially absorbed spectrum at energies larger than the
Lyman limit.  Consequently the HII region is reduced, and the X--ray
warming front is well behind the light radius.  This occurrence leads to a
larger absorption ring where the signal reaches $\simeq -20 \,
\mu$Jy in a 2 arcmin beam.

Imaging the gas surrounding a quasar in 21--cm emission could provide
a direct means of measuring intrinsic properties of the source, like
the emitted spectrum and the opening angle of quasar emission.  All
these features are within reach of the new generation radio telescopes
like {\sl SKA}.

\section{Conclusions}

The Wouthuysen-Field effect allows one to peer into the Dark Age.  The
observation of the neutral IGM in the redshifted 21--cm can give
insight into the thermal evolution of the diffuse hydrogen and thus
into the formation and evolution of radiation sources, at epochs when
the age of the universe is only $\simeq 0.3$ Gyrs.  In particular, the
epoch of the First Light can be seen as a deep ($\simeq 40$ mK)
absorption feature a few MHz wide against the CMB, at the
corresponding redshifted 21cm line.  Moreover, the density
perturbation field at a redshift $z\approx 5\div 20$ can be
reconstructed looking for mK fluctuations at $1$--$5$ arcmin
resolution in the radio sky, providing a determination of its
amplitude between the epoch probed by galaxy surveys and
recombination.  Finally, the first ionizing sources, like luminous
quasars, can be seen by identifying peculiar, ring--shaped signals
whose morphology depends on the source's age, luminosity and geometry.

\section*{References}



\begin{thebibliography}{99}

\bibitem{ref1} Madau, P., Meiksin, A., \& Rees, M. J. 1997, ApJ, 475, 429 (MMR)

\bibitem{ref2} Meiksin, A. 1999, invited review for the Square
Kilometre Array Radio Telescope Science Case, ed. R. Braun and
A. R. Taylor, astro-ph/9902384

\bibitem{ref3}  Tozzi, P., Madau, P., Meiksin, A., \& Rees, M. J. 1999,
ApJ submitted (TMMR)

\bibitem{ref4} Field, G.~B. 1958, Proc. I.R.E., 46, 240

\bibitem{ref5} Field, G.~B.  1959, ApJ, 129, 551

\bibitem{ref6} Couchman, H. M. P. 1985, MNRAS, 214, 137

\bibitem{ref7} Couchman, H.~M.~P., Thomas, P.~A., \& Pearce, F.~R. 1995, ApJ,
452, 797

\bibitem{ref8} Braun, R. 1998, Square Kilometer Array Radio Telescope
Science Case


\end{thebibliography}
\end{document}